\documentclass{pnastwo}

\newcommand{\bmb}[1]{\mathbf{#1}}

\newcommand{\citenum}[1]{\cite{#1}}

\url{}
\copyrightyear{}
\issuedate{Issue Date}
\volume{Volume}
\issuenumber{Issue Number}
\begin{document}

\title{The topological non-local  braid-group concept of information processing in brain, the different role of the gray and white matter}

\author{W. Jacak and J. Jacak \affil{1}{Wroc{\l}aw University of Science and Technology, Wroc{\l}aw, Poland}}

\contributor{Submitted to Proceedings of the National Academy of Sciences
of the United States of America}

%%%Newly updated.
%%% If significance statement need, then can use the below command otherwise just delete it.
\significancetext{Functioning of the neuron system in the brain is considered in terms of electrical signals in the web of axons and dendrites in the gray matter in where we suppose an information is encoding in the entangled web via activation of selected synapses.  We identify two types of physical mechanisms of neuro-signal transfer in the gray and white matter, correspondingly, which might be helpful in understanding of the role of both neuron fractions in information processing in mind utilizing the concept of topological-braid group nonlocal  type  of registration, storage and identification of information in the neuron web.   The  mechanism of the saltatory conduction in myelinated axons in the nerve system is described as the plasmon-polariton wave kinetics in ionic cytoplasm of axons with good agreement with experimental observations, precluding, however, the e-m response of myelinated axons. The plasmon-polariton  model elucidates the role of the myelin sheath in effective and fast  saltatory  communication, whereas the diffusive ion current kinetics in the gray matter serves to e-m resonance identification of in-entangled braids stored information of the memory.}

\maketitle

\begin{article}
\begin{abstract}
{The velocity of the action potential transduction along  myelinated  axons in the peripheral nervous system or in the white matter of  brain and spinal cord reaches hundreds of meters per second to assure proper functioning  of the body, which exceeds the ability of diffusive ion conduction.     We propose the new model of the saltatory conduction based on a plasmon-polariton kinetics in myelinated axons, which excludes, however, the white matter form the information storage and its identification in the brain via e-m response. We propose a nonlocal topological approach to information processing in the cortex of brain in consistence with the ion electricity of the gray matter and a supplementary only communication role of the white matter.}
\end{abstract}

\keywords{homotopy | braids |  plasmon-polariton | myelinated axon | saltatory conduction}

\abbreviations{PP, plasmon-polariton; e-m, electro-magnetic, EEG, electroencephalography, ac, alternating current}

\dropcap{T}he electric activity of nervous system is referred to transduction of action potential spike series along axons and dendrites of neuron cells connected together in electro-chemical synapses into a highly entangled web. Not all trajectories for signal travel are accessible  and the activation of selected synapses makes traces available to communication for longer or shorter time-periods depending on intensity and repetition of their activation. The classification of the resulted connection patterns may be done in topological terms of so-called pure braid groups \cite{birman} as will be demonstrated in the next paragraph. Elements of these groups display various inequivalent patterns of entanglement of treads (neuron filaments)  allowing, however, a precise  comparison and distinguishing of complicated  filament muddles in a mathematically precise  topological-homotopy way. Two braids are homotopic if one can  be transformed into another by continuous deformations without cutting.  For $N$ treads the number of various topologically inequivalent patterns of braidings is infinite though countable \cite{birman}, in practice limited only by size restrictions, like a finite diameter and length  of filaments. If one activates electrically such an entangled network pushing a charge current along its treads, one can notice that the unique pattern of  braiding   with a specific number of loops   might serve as multiloop e-m active circuit. The ac electric current in such braid-coil would generate thus   an individual also unique e-m local-space field distribution which could fall into resonance with the 
another web part representing the same braid group element located in the range of this dynamic field distribution. One can associate thus information messages (elements of the  memory)  imprinted in distinct elements of a pure braid  group   implemented in a  nonlocal manner  in fragments of the large neuron web of a brain via previous activation of some subset of synapses.  This massage could be temporary stored   in the neuron-braid corresponding to some mathematical braid from  the pure braid group  as long as the selected engaged synapses are still active (or are ready to be active). This information can be next invoked by e-m resonance with the similar braid  within the brain and temporary created in another its structure (e.g., in hippocampus) as the result of an influx of neuron signals from surroundings via senses. The new message tentatively imprinted in a temporary braid in the hippocampus might excite via e-m resonance the same braid  recorded in the past in the cortex, and in this way may cause identification of a new message in mind.  
To utilize the braids in the proposed  e-m recognition process,  the electrical ac currents are necessary to initiate a resonance response. The role of such ac-electricity may play brain waves which actually are  constantly  observed via EEG in the gray matter  of the cerebral cortex. For transduction of neuron-signal in non-myelinated axons and in all dendrites in the cortex, the diffusive  movement of ions in elongate and branched  neuron cells is employed. This electric-current-type communication is very well described by the cable theory  originated by W. Thompson (1854) \cite{kel} and widely applied to explanation of neural electro-physiology of dendrites and non-myelinated axons   (with various modifications \cite{cap,idi}). 

Hence, one can expect that the storage and identification 
of the information take place in the cerebral cortex taking an advantage of the huge information  capacity of  entanglement of neuron filaments creating various braid structures  enumerated by the pure braid group elements.  The topologically precise homotopy-resolution and a quick and complex  access to the stored messages  in entangled braids \cite{birman,mermin1979,ws,top}    vie e-m resonance utilizing additionally a few distinct frequencies of the  brain waves might be helpful to better understand the  functionality of the  memory and the pattern recognition by a mind. All these takes place in the gray matter of the cortex where the signal transduction is carried out by means of diffusive movement of ions equivalent to an ordinary electric ac current, which is able to induce local e-m  field in a coil. Different types of frequencies of brain waves might be here convenient to implement simultaneous independent  resonance channels for better  identification of multiloop coils created by braids \cite {top}.

\section{Braid groups to store and proceed information in gray  matter}

The braid groups have been introduced to describe entangled bundles of classical trajectories of $N$ particle systems \cite{birman}. There are investigated and applied two  types of braid groups, the first one with assumption of indistinguishability of identical particles, called as the full braid groups and applied to characterize quantum statistics of particles \cite{ws} and the second type---the so called pure braid groups for ordinary distinguishable particles \cite{birman,ws}. The latter are suitable to describe and classify highly entangled multi-filament webs. Both full and pure braid groups are infinite for $N$ particles moving on the plane but are finite groups when particles move in 3D space. In 3D there is room to disentangle lines representing particle trajectories, what is, however, impossible in 2D (in 2D particle must twist up one around another and cannot hop beyond or beneath others). The pure braid group of $N$ distinguishable particles on a plane may be presented as the set of  $N$ treads in 3D with their ends fixed to steady points  but braided on the way in an arbitrary manner. Thus the 3D web of entangled treads as of neurons in a brain can be mapped on 2D pure braid group.  The patterns of such braidings are  unique and cannot be transferred one into another one without cuttings of treads and can be precisely enumerated  by the pure braid group elements. If one imagines the $N$ end and $N$ final points of treads steady fixed in 3D space, then the  structure of linking them braids of threads---the 2D $N$-element pure braid group elements  reproduce one-to-one
3D webs of arbitrary entangled nets of filaments, like a webs of neurons linked via synapses in the brain.

\subsection{Pure braid group}

\begin{figure*}
\centering
\scalebox{1.2}{\includegraphics{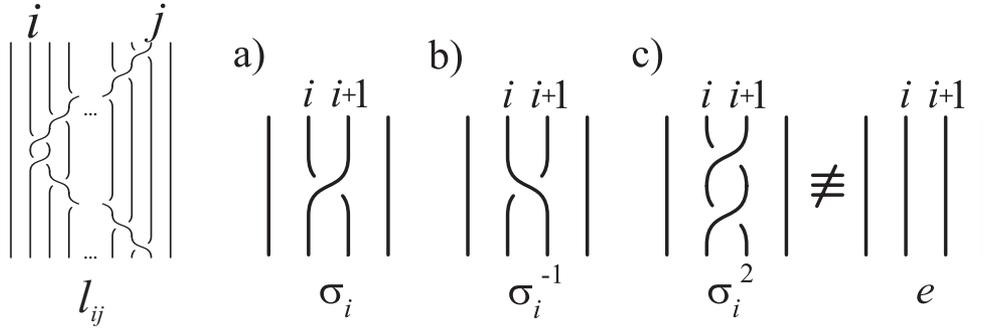}}
\caption{\label{generator} Generator $l_{ij}$ of the pure braid group (left), the exchange of neighboring particles, $\sigma_i$ (a) and its inverse, $\sigma_i^{-1}$ (b) and the reason why the braids on a plane are complicated (c).}
\end{figure*}

The topology characteristics of entangled filament webs, like neuron webs, may be described in terms of homotopy reflecting the complexity of trajectories of many particle systems \cite{spanier1966}. The first homotopy group of the space $D$ marked with $\pi_1(D)$, is a collection of topologically non-equivalent (non-homotopic, i.e.,  one cannot be continuously deformed to another) classes of closed trajectories in the space $D$. If this space is a configuration space of a system of $N$   particles, each of which runs its own trajectory on a manifold $M$, then the appropriate $\pi_1$ homotopy group is called a braid group. The configuration space for $N$ identical particles located on $M$ manifold (e.g.,  $R^n$) is defined as follows,
$F_N(M)=M^N\setminus\Delta$,
for distinguishable identical particles;
$M^N$ is a $N$-times normal product of $M$ manifold, $\Delta$ is a set of diagonal points (where the coordinates of two or more particles coincide), which has to be subtracted in order to conserve the number of particles in the system (particles cannot hide one behind another). 
The braid group is the first homotopy group \cite{spanier1966,mermin1979} $\pi_1$ of the configuration space of a system of $N$ particles. Braid groups represent available trajectories of a system of $N$ particles (while  are not referring to any specific dynamics).
The pure braid group which we will utilize to neuron web is thus defined as
\begin{equation}
\pi_1(F_N(M))=\pi_1(M^N\setminus\Delta ).
\end{equation}
When braiding trajectories,  the initial ordering of particles must be unchanged. By $l_{ij}$ the elementary generators of the pure braid group can be assigned 
(Fig. \ref{generator}) \cite{birman}, they correspond to the simplest entanglement  of two trajectories of a  particle  pair, $i$-th and $j$-th, while the rest of particle trajectories remain not-tangled, and can be presented with the use of  $\sigma_i$ (exchanges of neighboring particles $i$-th and $i+1$-th  not conserving their positions \cite{birman}),
\begin{equation}
l_{ij}= \sigma_{j-1} \cdot \sigma_{j-2}....\sigma_{i+1}\cdot \sigma_i^2\cdot
\sigma_{i+1}^{-1}...\sigma_{j-2}^{-1}\cdot \sigma_{j-1}^{-1},\;\; 1\geq i\geq j\geq N-1.
\end{equation}
Pure braid group can be considered as an abstract mathematical group generated by $l_{ij}$ defined by the following relations \cite{birman,ws},
\begin{equation}
l_{rs}^{-1} \cdot l_{ij} \cdot l_{rs} =\left\{\begin{array}{ll} 
{l_{ij},} & {i<r<s<j} 
\\ {l_{ij},} & {r<s<i<j} 
\\ {l_{rj} \cdot l_{ij} \cdot l_{rj}^{-1},} & {r<i=s<j} 
\\ {l_{rj} \cdot l_{sj} \cdot l_{ij} \cdot l_{sj}^{-1} \cdot l_{rj}^{-1} ,} & {i=r<s<j} 
\\ {l_{rj} \cdot l_{sj} \cdot l_{rj}^{-1} \cdot l_{sj}^{-1} \cdot l_{ij} \cdot l_{sj} \cdot l_{rj} \cdot l_{sj}^{-1} \cdot l_{rj}^{-1} ,} & {r<i<s<j}. \end{array}\right.
\end{equation}

We propose to identify a fragment of the neuron web connected via active synapses predefined by a prior record  of some information in filament web representing  some pattern from the pure braid group. We have proved that one can use the binary code to record in 3-filament bundle  of any message (two generators of the corresponding  $N=3$ pure braid group serve to code 0 and 1 bits and the third generator encodes the end of a word). We have solved also the problem of coding of $N$-letter alphabet message in $N$-order pure braid group \cite{top}. We demonstrated that the ratio of an information efficiency of $N$-order pure braid group  to resources needed  to organize a physical web (like a neuron web with $N$ entangled filaments) attains its maximum at $N=20-30$, which agrees with the number of phones in majority languages \cite{ws}. This might suggest that words are codded in relatively small number of filaments in entangled bunches.  

In the braid scenario we propose to code information in  entanglement patterns in pieces of the neuron web in the brain selected by activation of some synapses. This is a topological nonlocal type of coding and is more efficient that the conventional addressing of synapse registers ($N$ synapses addressing offers only $N!$ various connections with another $N$ synapse register, whereas the number of distinct brads of threads linking this registers is infinite, limited only by a physical constraint of finite size (diameter and length) of threads).

To identify neuron braids we propose the e-m resonance utilizing ion currents which flow in neuron filaments of the gray matter. The different role plays the white matter. The  effective and fast transduction of the action potential through long myelinated  axons of the peripheral nerves and in the white matter of the central nervous system undeniably stands behind the motorics and other communication functions of the body and requires at least 100 times faster signaling than that one offered by diffusion-type ion flows upon the cable theory. Actually such a high-speed signaling in meylinated axons is observed and called as the saltatory conduction of action potential \cite{lil}. 
In myelinated axons the action potential occurs in so-called  Ranvier nodes \cite{fr} and is renewing on subsequent series of these nodes upon a very well recognized mechanism of Hodgkin--Huxley membrane model \cite{fr}. However, in-between the Ranvier nodes along the myelinated sectors of an axon the  signal hops,  which accelerates  strongly the action signal transduction along the periodically myelinated axon structure. The velocity of this saltatory conduction is at least  two order faster than the diffusion ion current. Such an acceleration  is absolutely required for proper  functionality of long distance signaling   in peripheral nervous system but also for quick communication in the brain and in spinal cord in their white matter. It is interesting that in the brain the white matter is located in inner structures beneath the cortex, but in the spinal cord inversely. This might be linked with a different role of the gray matter in the cerebral cortex and in the spinal cord---the cortex is the place of memory storage and of it comparison or identification with new messages from surroundings accessing a brain via senses, whereas the gray matter in the spinal cord serves to identification of internal body informations necessary to the control of body functioning. The latter must be precise and error-less thus the gray matter in a spinal cord ought to be isolated from surroundings as best as possible, here  by an outer   white matter layer. It has been  demonstrated  that by e-m external field, especially its magnetic component, one can influence and  trans-cranial perturb identification process in the cortex,  what would be, however, very danger in the spinal cord---thus the latter must be better isolated than the cortex in brain.  

 The white matter take part thus in communication but is not utilized to recognition of braids with stored informations which is the role of the gray matter. Actually the signal transduction in myelinated axons is not of electrical current character and thus cannot take part in e-m identification of braids. 

The above described scenario  is consistent with the plasmon-wave type character of the saltatory conduction in myelinated axons \cite{jacak2015}. The saltatory conduction is apparently of a wave type because it is maintained even if a myelinated axon is divided into two disjoint pieces with ends separated slightly by an insulating barrier impossible to be crossed by any diffusive ion current.

In the present paper we describe a  model for the saltatory conduction based on kinetic properties of collective plasmon-polariton (PP) modes propagating along linear and periodically myelinated electrolyte system, which in the case of axon is the thin cord of the nerve cell periodically wrapped by Schwann cells creating thick myelin sheath (Schwann cells myelinate axons in the peripheral system, whereas in the central neural system the white matter is built from axons myelinated by the oligodendrocyte cells). PPs  were previously investigated and  understood based on the very well developed domain of plasmonics \cite{plasmons,pitarke2006}. There was   demonstrated  a long range low-damped propagation of PPs along metallic nano-chains \cite{atwater2005}. The main properties of these collective excitations occurring in metallic nano-chains  on the conductor/insulator interface \cite{maradudin,deabajo} by hybridization of  surface plasmons (i.e.,  fluctuations of charge density of electron fluid  on the metal surface) with the electro-magnetic wave, are as follows: \begin{enumerate}
    \item much lower velocity of PPs in comparison to the light velocity, which gives much shorter wave-length of PPs than light wave-length with the same frequency, this causes a strong discrepancy of momenta of PPs and photons with the same energy and PPs do not interact with e-m waves, i.e., photons cannot be excited or absorbed by PPs,
\item radiative losses of PPs do not exist   and PP attenuation is only due to Ohmic losses, which makes metallic nano-chains almost perfect  waveguides for PPs,
\item long range and practically undamped propagation of PPs is observed experimentally \cite{atwater2005,atwater2003b}.
\end{enumerate} PPs in metallic nanostructures are regarded as prospective for future applications in opto-electronics where conversion of the light signal onto PP signal allows for avoidance of the diffraction constraints strongly limiting miniaturization of conventional opto-electronic devices (as the nano  scale of electron confinement  conflicts  with the several orders larger scale of the wave-length of light with energy similar to that one of electrons trapped in the nano-scale) \cite{deabajo,citrin2005}. 

Some of the unique properties of PPs in metallic nano-chains are of especial  interest  as  they can be repeated in periodic linear alignments of electrolyte systems with ions instead of electrons as charge carriers. Due to larger mass of ions in comparison to electrons and lower concentration of ions in  electrolytes than the concentration  of electrons in metals,  plasmon resonances in finite ionic systems (e.g., in liquid electrolyte confined to the finite volume by appropriately formed membranes frequent in biological cell organization) occur in a different than in  metals scale of micrometers instead of nanometers  and for several orders lower frequency/energy (depending on ion concentration). 

For spherical electrolyte micro-system one deals with the surface and volume plasmons in analogy to metallic nanosphere \cite{jacak2014a,jacak5}. The ionic surface plasmon frequencies are given for $l$th multipole  mode  by the formula $\omega_l=\omega_p \sqrt{\frac{l}{\varepsilon(2l+1)}}$, where the bulk plasmon frequency $\omega_p=\sqrt{\frac{4 \pi q^2 n}{m}}$ ($n$ is ion concentration, $q$ and $m$ are ion charge and mass, respectively, $\varepsilon$ is the dielectric relative  permittivity of the surroundings). For  dipole-type surface plasmons ($l=1$) the frequency is given by so-called Mie-type formula, \cite{mie,jacak5} $\omega_1=\frac{\omega_p}{\sqrt{3 \varepsilon}}$. 

Plasmon oscillations in a single micro-sphere of electrolyte  intensively irradiate their own energy and are quickly damped due to Lorentz friction losses (i.e., due to  irradiation of the electro-magnetic  wave by oscillating charges \cite{lan,jackson1998}) which for large systems with large number of ions participating in plasmon oscillations (thus strengthening the Lorentz friction) are  much greater than the Joule-heat dissipation caused by Ohmic losses due to carriers scattering (collisions of ions with other ions, solvent and admixtures  atoms and with boundary of the system), as illustrated in Fig. \ref{lorentz45}, where the crossover in the size dependence of ion plasmon damping rate is presented.

Surprisingly, in the linear chain of  ionic systems the irradiation losses are completely reduced to zero, exactly in the same manner as in metallic nano-chains \cite{citrin06,markel2007,jacak2013}. The large irradiation energy  losses expressed by the Lorentz friction at each chain element  \cite{lan,jackson1998} are ideally compensated by the income of energy due to radiation of the chain rest. In the result the radiative losses are  ideally balanced by mutual radiation exchange and there remains only relatively small Ohmic irreversible energy dissipation. The collective PPs can thus propagate in the ionic chain with strongly reduced damping  as in almost perfect waveguide. If additionally the energy is supplemented  in a minor scale to compensate  small  Ohmic losses, the propagation can reach arbitrarily long distances  without any damping. This is what is observed in saltatory conduction in myelinated axons, moreover,  PPs in this axons cannot be neither detected not perturbed by external e-m field.

\section{PP propagation in linear periodic  ionic system modeling a myelinated axon}

In more detail the interaction  between the ion chain elements  can be regarded as a dipole-type coupling \cite{atwater2003b}. For chains of ionic spheres, one can adopt the results of the corresponding analysis for metallic chains, which supports a dipole model of interaction between spheres
\cite{citrin2005,schatz2003,schatz2004}. 

The dipole interaction resolves itself to the electric and magnetic fields created by an oscillating dipole $\mathbf{D}(\mathbf{r},t)$ at any distant point. If this point is represented by the vector $\mathbf{r}_0$ (with one end fixed at the end of $\mathbf{r}$ where the dipole is placed), then the electric field produced by the dipole $\mathbf{D}(\mathbf{r},t)$ has the following form, including the  retardation \cite{lan,jackson1998}:
\begin{equation}
\label{electric}
\begin{array}{l}
 \mathbf{E} (\mathbf{r},\mathbf{r_0},t)=\frac{1}{\varepsilon}
\left(-\frac{\partial^2}{v^2\partial t^2} \frac{1}{r_0} -\frac{\partial}{v \partial t} \frac{1}{r_0^2}-\frac{1}{r_0^3}\right)\mathbf{D}(\mathbf{r},t-r_0/v)\\ 
+ \frac{1}{\varepsilon}
\left( \frac{\partial^2}{v^2\partial t^2}\frac{1}{r_0}+\frac{\partial }{v \partial t} \frac{3}{r_0^2}+\frac{3}{r_0^3}\right) \mathbf{n}_0(\mathbf{n}_0\cdot \mathbf{D}(\mathbf{r}, t-r_0/v)),\\
\end{array}  
\end{equation}   
with 
$\bmb{n}_0=\frac{\bmb{r}_0}{r_0}$ and   
$v=\frac{c}{\sqrt{\varepsilon}}$, i.e., the light velocity in a dielectric medium.
The terms with denominators of  $r_0^3$, $r_0^2$, and $r_0$ usually are  referred to as the near-field, medium-field, and far-field components of the interaction, respectively.
The above formula allows for the description of the mutual interaction of the plasmon dipoles at all spheres in the chain.
The spheres in the chain are numbered by integers $l$ and 
 the equation for the surface plasmon oscillation of the $l$th sphere can be written as follows ($d$ denotes separation between subsequent sphere centers) \cite{jacak5,jacak2015}:
\begin{equation}
\label{tra111111}
\begin{array}{l}
 \left[\frac{\partial^2}{\partial t^2}+  \frac{2}{\tau_0}  \frac{\partial}{\partial t} +\omega_1^2\right]D_{\alpha}(ld,t)\\
 =\varepsilon \omega_1^2a^3
  \sum\limits_{m=-\infty, \;m\ne l }^{m=\infty} 
E_{\alpha}\left(md,t-\frac{|l-m|d}{v}\right)\\
+\varepsilon \omega_1^2a^3 E_{L\alpha}(ld,t) +\varepsilon \omega_1^2a^3 E_{\alpha}(ld,t),\\
  \end{array}
\end{equation}
$\alpha=z$ indicates the longitudinal polarization of dipole oscillations, whereas $\alpha=x(y)$  the transverse polarization (the chain orientation is assumed to be along the $z$ direction).
The first term on the right-hand side of Eq. (\ref{tra111111}) describes the dipole coupling between spheres, and the other two terms correspond to the plasmon attenuation due to the Lorentz friction irradiation losses \cite{jacak2015,lan,jackson1998}  and the driven field arising from an external electric field, respectively; $\omega_1=\frac{\omega_p}{\sqrt{3\varepsilon}}$ is the frequency of the dipole surface plasmons \cite{jacak2014a}. Ohmic losses are included via the term $\frac{2}{\tau_0}$ similarly as  in metals \cite{atwater2003} but with the Fermi velocity of electrons substituted by the mean velocity of ions according to the Boltzmann distribution regardless of quantum statistics of ions (unlike electrons in metals ions are not quantumly degenerated and are described by classical Boltzmann distribution),
\begin{equation}
\label{form}
\frac{1}{\tau_0}=\frac{\mathsf{v}}{2\lambda_B}+\frac{C \mathsf{v}}{2a},
\end{equation}
where $\lambda_B$ is the mean free path of carriers in  bulk electrolyte, $\mathsf{v}$ is the mean velocity of ions at a temperature $T$, $\mathsf{v}=\sqrt{\frac{3kT}{m}}$, $m$ is the mass of the ion, $k$ is the Boltzmann constant, $C$ is a constant of order unity (to account for the type of scattering of carriers by the system boundary) and $a$ is the radius of a electrolyte sphere. The first term in the expression for $\frac{1}{\tau_0}$ approximates ion scattering losses of the same character  as those occurring in a bulk electrolyte (collisions with other ions, with solvent and admixture  atoms), whereas the second term describes the losses due to scattering of ions on the boundary of a  sphere of radius $a$.  According to Eq. (\ref{electric}) ($\mathbf{n}_0$ is oriented along the chain, i.e., the $z$-axis), we can write the following quantities that appear in Eq. (\ref{tra111111}):
\begin{equation}
\begin{array}{ll}
E_z(md,t)&=\frac{2}{\varepsilon d^3}
\left(\frac{1}{|m-l|^3}+\frac{d}{v|m-l|^2}\frac{\partial}{\partial t}\right)\\
&\times D_z(md,t-|m-l|d/v),\\
E_{x(y)}(md,t)&=-\frac{1}{\varepsilon d^3}\left(\frac{1}{|m-l|^3}+\frac{d}{v|m-l|^2}\frac{\partial}{\partial t}+
\frac{d^2}{v^2 |l-d|}\frac{\partial^2}{\partial t^2} \right)\\
&\times D_{x(y)}(md, t-|m-l|d/v).\\
\end{array}
\end{equation}

Because of the periodicity of the chain one can assume a wave-type collective solution of the dynamical equation in the form of the Fourier component  for solution of Eq. (\ref{tra111111}):
\begin{equation}
\label{eq8}
\begin{array}{l}
D_{\alpha}\left(ld,t\right)=D_{\alpha}\left(k,t\right)e^{-ikld},\\
0\leq k \leq\frac{2\pi}{d}.
\end{array}
\end{equation}
In the Fourier picture of Eq. 
(\ref{tra111111}) (the discrete Fourier transform (DFT) with respect to positions and the continuous Fourier transform (CFT) with respect to time), this solution takes a form similar to that of the solution for the case of phonons in 1D crystals. 
Let us note that the DFT is defined for a finite set of numbers; therefore, we consider a chain with $2N+1$ spheres, i.e., a chain of finite length $L= 2Nd$. Then, for any discrete characteristic $f(l),\;\;l=-N,...,0,...,N$ of the chain, such as a selected polarization of the dipole distribution in the PP wave along the chain, one must consider the DFT picture $f(k)=\sum\limits_{l=-N}^{N}f(l)e^{ikld}$ where $k=\frac{2\pi}{2Nd}n,\;n=0,...,2N$. This means that $kd\in[0,2\pi)$ because of the periodicity of the equidistant chain. The usual Born-Karman periodic boundary condition, $f(l+L)=f(l)$, is imposed on the entire system, resulting in the form of $k$ given above. For a chain of infinite length, one can take the limit $N\rightarrow \infty $, which causes the variable $k$ to become quasi-continuous, although $kd\in[0,2\pi)$ still holds.    

The Fourier representation of Eq. (\ref{tra111111}) has the following form:
\begin{equation}
\label{aaa1}
\begin{array}{l}
\left(-\omega^2-i\frac{2}{\tau_0}\omega +\omega^2_1\right)D_{\alpha}(k,\omega)\\
=\omega_1^2\frac{a^3}{d^3}F_{\alpha}(k,\omega)D_{\alpha}(k,\omega)+
\varepsilon a^3 \omega_1^2 E_{0\alpha}(k,\omega),\\
\end{array}
\end{equation}
with 
\begin{equation}
\label{aaapodluzneipoprzeczne}
\begin{array}{l}
F_z(k,\omega)=4\sum\limits_{m=1}^\infty \left(\frac{cos(mkd)}{m^3}cos(m\omega d/v)\right.\\
\left. +\omega d /v \frac{cos(mkd)}{m^2}sin(m\omega d/v)\right)\\
+2i \left[\frac{1}{3}(\omega d /v)^3+2\sum\limits_{m=1}^\infty \left(\frac{cos(mkd)}{m^3}sin(m\omega d/v)\right.\right.\\
\left.\left. -\omega d/v\frac{cos(mkd)}{m^2}cos(m\omega d/v)\right)\right],\\
F_{x(y)}(k,\omega)=-2\sum\limits_{m=1}^\infty \left(\frac{cos(mkd)}{m^3}cos(m\omega d/v)\right.\\
\left.+\omega d /v \frac{cos(mkd)}{m^2}sin(m\omega d/v)
 -(\omega d/v)^2\frac{cos(mkd)}{m}cos(m\omega d/v)\right)\\
-i \left[-\frac{2}{3}(\omega d /v)^3+2\sum\limits_{m=1}^\infty \left(\frac{cos(mkd)}{m^3}sin(m\omega d/v)\right.\right.\\
\left.\left.+\omega d/v\frac{cos(mkd)}{m^2}cos(m\omega d/v)\right.\right.\\
\left.\left. -(\omega d/v)^2\frac{cos(mkd)}{m}sin(m\omega d/v)\right)\right].\\
\end{array}
\end{equation}
We see that $Im F_{\alpha}(k,\omega)\equiv 0$ within the light cone, which indicates the perfect quenching of the irradiation losses at each sphere in the chain (it means that to each sphere the same amount of energy incomes from other spheres as  outflows due to intensive Lorentz friction). One can easily  verify this property  as the related infinite sums in Eq. (\ref{aaapodluzneipoprzeczne}) can be found analytically \cite{gradst}. Eq. (\ref{aaa1}) is highly nonlinear  with respect to the complex $\omega$ and can be solved both perturbatively in the analytical manner \cite{jacak2013} or numerically--more accurately  \cite{jacak2014}. The determined solutions for $Re\omega$, $Im\omega$ (i.e., for self-frequency and damping of PP, respectively) can be applied to the model of axon as the effective chain of electrolyte spheres with  ion concentration   adjusted to real neuron parameters.

 The resonance frequency $Re\omega(k)$, its $k$-derivative being the  group velocity of a particular PP mode  and the attenuation rate $Im\omega(k)$ of dipole PP modes numbered by the wave vector $k$, derived by solution of  Eq. (\ref{aaa1}) \cite{jacak2013,jacak2014,jacak2015}  for ionic chains  with  exemplary concentration, chain size  and ion parameters are plotted in Fig. \ref{figk1}.

\section{Fitting  plasmon polariton kinetics to  axon parameters}

\label{axon}

For the bulk ion plasmon frequency $\omega_p=\sqrt{\frac{q^2 n 4 \pi}{m}}$ (with  assumed for a model the ion charge  $q=1.6 \times 10^{-19}$ C  and the ion mass $m=10^4 m_e$, where $m_e=9.1 \times 10^{-31}$ kg is the mass of an electron) and for $n=2.1 \times 10^{14}$ 1/m$^3$ ion concentration we obtain  the Mie-type  frequency for ionic dipole oscillations, $\omega_1
\simeq 0.1 \frac{\omega_p}{\sqrt{3 \varepsilon_1}}\simeq 4 \times 10^6$ 1/s, where the relative permittivity of water is $\varepsilon_1 \simeq 80$  for frequencies in the MHz range \cite{epsilon} (although for higher frequencies, beginning at approximately 10 GHz, this value decreases to approximately 1.7, corresponding to the optical refractive index of water, $\eta\simeq \sqrt{\varepsilon_1}=1.33$). Let us emphasize that the axon consists of a cord with a small diameter of $2r$, and this thin cord is wrapped with a myelin sheath of a length of $2a$ per segment between two Ranvier nodes; however, for the effective model, we consider fictitious  electrolyte spheres of radius $a$. Thus, the auxiliary concentration $ n$ of ions in the fictitious spheres  corresponds to a concentration of ions in the cord of $n'=\frac{n 4/3 \pi a^3}{2a \pi r^2}$, which yields a typical concentration of ions in a nerve cell of $n' \sim 10 $ mM (i.e., $\sim 6\times 10^{24}$ 1/m$^3$). This is because all ions participating in the dipole oscillation  correspond in the sphere model to a much smaller volume in the real system, that of the thin cord portion (the insulating myelinated sheath consists of a lipid matter without any ions). The insulating, relatively thick myelin coverage creates the periodically broken channel required for PP formation and its propagation. To reduce the coupling with the surrounding inter-cellular electrolyte and protect against any leakage of the PPs, the myelin sheath must be sufficiently thick, much thicker than what is required merely for electrical insulation. Moreover, to accommodate the conductivity parameters which we have calculated for a spherical geometry to the highly prolate geometry of the real oscillating ionic system, we have accounted for the fact that the Mie-type frequency of the longitudinal oscillations in long thin rod  must be lower than that one for a sphere with a diameter equaled to the elongation axis. As a rough estimate, we assumed a correction factor of 0.1 consistent with the similar aspect-reducing factor in metallic nanoparticles \cite{elipsa1,elipsa2}.

For the resulting Mie-type frequency, $\omega_1\simeq 4 \times 10^6$ 1/s, one can determine the PP self-frequencies in a chain of spheres of radius 50 $\mu$m (for a Schwann cell length of $2a=100\;\mu$m) and for small chain separations  $d/a=2.01$, $2.1$, and $2.2$ (corresponding to Ranvier node lengths of $0.5$, $5$, and $10\;\mu$m, respectively) within the approach presented above (via solution of Eq. (\ref{aaa1})). The derivative of the obtained self-frequency with respect to the wave vector $k$ determines the group velocity of the PP modes. The results are presented in Fig. \ref{axon-100}. We notice that for the ionic system parameters listed above, the group velocity of the PPs reaches 100 m/s for the longitudinal mode with polarization suitable to the prolate geometry  assuming that the initial post-synaptic action potential or that from the synapse hillock excites predominantly longitudinal ion oscillations.

Note that for $\omega_1=4\times 10^6$ 1/s and $a=50 \;\mu$m, the  known in metallic nano-chains  light-cone interference conditions \cite{jacak2014} $kd-\omega_1 d/c=0$ and $kd+\omega_1 d/c=2\pi$ are fulfilled for extremely small values of $kd$ and $2\pi -kd$, respectively (of the order of $10^{-6}$ for $d/a\in[2,2.5]$), and thus are negligible for PP kinetics under these conditions (though were important for metals \cite{jacak2014,markel2007}). The related singularities on light-cone induced by the far- and medium-field contributions to the dipole interaction \cite{jacak2014,markel2007} are pushed to borders of the $k$ wave vector domain and thus are unimportant in the case of considered ionic system. Hence,  the quenching of the radiative losses (i.e., the perfect compensation of the Lorentz friction in each sphere  by the radiation income from other spheres in the chain)  for the PP modes in the axon model occurs throughout practically the entire $kd\in[0,2\pi)$ region. Additionally, the above-mentioned singularities \cite{jacak2014} are characteristic of infinite chains and therefore cannot fully develop because the nerve model electrolyte  chains are of a finite length, whereas other effects, as quenching of irradiation losses, occur for finite chains due to very fast convergence of sums in Eq. (\ref{aaapodluzneipoprzeczne}) with denominators $m^2$ and $m^3$ (practically, the chain consisting of only 10 elements  exhibits almost same properties as the infinite one, except for singularities  \cite{citrin2005,jacak2013}).

Although the ionic system chain model for a myelinated axon appears to be a crude approximation of the real axon structure, it can serve for the comparison of the energy and related time scales of PP propagation implied by the model with the observed kinetic parameters of nerve signals.  
In the model, the propagation along the axon  of a PP that is excited by an initial action potential on the first Ranvier node (next to  the synapse or, for the reverse signal direction, in the synapse hillock) sequentially ignites consecutive Ranvier node blocks of $Na^+$ and $K^+$ ion gates, and the resulting firing of the action potential traverses the axon with the velocity  approximately 100 m/s, consistent with the velocity actually observed in myelinated axons. The PP ignition  of consecutive Ranvier nodes releases creation of the same activation potential pattern aided by the external   energy supply  at each  Ranvier node block. Because of the nonlinearity of this functional across-membrane-ion-channel-block, the signal growth saturates at a constant level, and the overall timing of each action potential spike has the stable shape of the local polarization/depolarization scheme in the short unmyelineted fragment of the axon cell membrane that corresponds to the Ranvier node. The permanent supply of energy associated to creation of the action potential spikes at one-by-one firing Ranvier nodes contributes also to the PP dynamics assuring the  amplitude of each Ranvier node dipole excitation being beyond the activation threshold. The external  energy supply (ATP/ADP cell mechanism) assisting the action potential forming at Ranvier nodes residually compensates  the Ohmic  losses of PP  mode propagating along the axon  and ensures undamped its propagation over an unlimited range. Although the entire signal cycle of the action potential spike forming  on a single Ranvier node block requires several milliseconds (or even longer when one includes the time required to restore the steady-state conditions, which, on the other hand, conveniently locks the reversal of the firing signal), subsequent nodes are ignited more rapidly than this period, corresponding to the velocity of the PP wave packet triggering ignition of consecutive Ranvier nodes, as illustrated in Fig. \ref{fig666666}. The direction of the velocity of the PP wave packet is adjusted to the semi-infinite geometry of the  chain (in fact, the chain is finite and is excited at one end) and protected by inertia  of Ranvier node action potential blocks. The firing of the action potential that is triggered by the PP  traverses along the axon in only one direction because the nodes that have already fired have had their $Na^+$ gates inactivated and will require a relatively long time to restore these gates to their original status (the entire sodium/potassium block sequence requires time of the order of even a second  as well as a sufficient energy supply to bring the concentrations to their normal values via cross-membrane active ion pumps against concentration gradient). The described above PP scheme for the ignition of the action potential spikes in the chain of Ranvier nodes along the axon is thus well consistent with the saltatory conduction observed in myelinated axons. An observation that firing of the action potential can move in two opposite directions simultaneously if one ignites some central Ranvier node of a passive axon,  as well as an observation of maintaining  of the firing traverse despite breaks of the axon cord or damage of even few Ranvier nodes,  agree well with the collective wave-type PP model of the saltatory conduction, in contrast to lack of a satisfactory explanation in models based on the cable theory or within the compression-soliton model.

In Fig. \ref{axon-predkosc}, the group velocity for an action potential traversing a firing myelinated axon is plotted for various diameters of the internal cord of the axon, with a length of $100\;\mu$m for each myelinated sector wrapped by Schwann cells, and for Ranvier intervals of 0.5 $\mu$m, 5 $\mu$m, and 10 $\mu$m. The dependence on the length of the Ranvier interval is weak (i.e., negligible at the considered scale), but the increase in the velocity with increasing diameter  of the internal cord is significant, similarly as in the real axons with growing diameter.

To comment on the appropriateness of the chain model to the axon case let us note that even though the thin core of the axon is a continuous electrolyte conducting  fiber, the surface electro-magnetic field can be closely pinned to the linear conductor similarly as for the Goubau line \cite{goubau,sommerfeld} and if the inner cord is periodically wrapped by a dielectric shell, the plasmon resonantly coupled with the e-m field propagates as the strongly concentrated PP along the chain of periodic segments despite the continuity of the inner fiber. The fragments of the thin axon cord wrapped with the myelin thick shells with typical for neurons  ion concentration $n'\simeq 10 $ mM $\simeq 6\times 10^{24} $ 1/m$^3$ inside the cell, can be equivalently modeled by  spheres with diameter equaled to the length of cord fragment and with the ion concentration $n\simeq 2 \times 10^{14} $ 1/m$^3$ (for the axon cord diameter assumed 100 nm) assuring  the same  number of ions in the segment taking part in dipole oscillations. Such a model is justified by the same structure of the dynamics equation for dipole plasmon fluctuations in the chain of spheres, i.e., of Eq. (\ref{tra111111}), and of its modification for the prolate spheroid or elongate cylindrical rod chain. 
This modification resolves to the substitution in  Eq. (\ref{tra111111}) of isotropic $\omega_1$ frequency by frequencies different for each polarization $\omega_{\alpha 1}$, i.e.,    
\begin{equation}
\begin{array}{l}
 \left[\frac{\partial^2}{\partial t^2}+  \frac{2}{\tau_{\alpha 0}}  \frac{\partial}{\partial t} +\omega_{\alpha 1}^2\right]D_{\alpha}(ld,t)\\
 ={\cal{A}}
  \sum\limits_{m=-\infty, \;m\ne l }^{m=\infty} 
E_{\alpha}\left(md,t-\frac{|l-m|d}{v}\right)\\
+{\cal{A}} E_{L\alpha}(ld,t) +{\cal{A}} E_{\alpha}(ld,t),\\
\end{array}
\end{equation}
where ${\cal{A}}= V \frac{n q^2}{m}$  is a shape independent factor proportional to the number of ion carriers with concentration $n$ in the volume of the spheroid with semi-axes $a,b,c$, $V=\frac{4\pi}{3}abc = \frac{4 \pi}{3}a^3$ (the latter for a sphere). Taking into account that the plasmon frequency in bulk electrolyte with ion concentration $n$ equals to  $\omega_p=\sqrt{\frac{n q^2 4 \pi}{m}}$, one can rewrite ${\cal{A}}$ as follows, ${\cal{A}}= \frac{abc \omega_p^2}{3}= \varepsilon a^3 \omega_1^2$ (the latter for a sphere, for which $\omega_1=\frac{\omega_p}{\sqrt{3\varepsilon}}$).
Ohmic losses can be  included via the anisotropic term,
$\frac{1}{\tau_{\alpha 0}}=\frac{\mathsf{v}}{2\lambda_B}+\frac{C \mathsf{v}}{2a^{\alpha}},$
where  $a^{\alpha}$ is the dimension (semi-axis) of the spheroid in the direction $\alpha$ (equaled to $a,b,c$ for a spheroid). The first isotropic term in the expression for $\frac{1}{\tau_{\alpha 0}}$ approximates ion scattering losses such as those occurring in the bulk electrolyte (thus is isotropic), whereas the second  term describes the losses due to the scattering of ions on the anisotropic boundary of the spheroid. 
The dipole coupling is independent of the shape of chain elements.
The described above mutual independence of dipole oscillations with distinct  polarizations follows from the linearity of the dynamical equation with respect to the dipole strength, regardless of the metal or electrolyte conducting elements.

Because the structure of the dynamics equation  is not affected by the anisotropy, thus its   solutions for each polarization have the same form as in the spherical case with exception for the modification of the related frequency of dipole oscillation in each direction and for the small correction to the orientation dependent contribution to the scattering ratio (this part being related to the boundary scattering of carriers). 
One can thus renormalize the equation for dipole oscillations independently for each polarization direction introducing in a phenomenological manner the resonance oscillation frequency for each direction $\omega_{\alpha 1}$ (these can be estimated numerically, whereas for a sphere $\omega_1=\frac{\omega_p}{\sqrt{3\varepsilon}}$; in general, the longer semi-axis the lower the related dipole oscillation frequency is). Moreover, the   size-correction of the boundary scattering term must be taken into account, as described above (favorably reducing longitudinal PP damping for prolate geometry).
Such a renormalization was accounted for in the considered model of axon segment chain.

The periodic structure of the myelinated axon actually does not form a chain of electrolyte spheres, but rather is a thin electrolyte cord wrapped  in myelin sheath on periodically distributed sectors separated by very short unmyelinated intervals---Ranvier nodes. This is, however, not important for PP propagation because PP is hybrid of dipole plasmon oscillations with the electro-magnetic wave. The periodic structure of the dielectric isolation even without fragmentation of inside cord  allows  for  consideration of collective  PP wave-type oscillations propagating along the whole system  which resolves itself to  synchronic of wave-type  dipole polarization of consecutive Ranvier nodes. Taking into account that Ranvier nodes are very short,  these polarization dipoles  are equivalent with similarly synchronically oscillating  dipoles of myelinated sectors. The latter   can be thus treated as dipole surface plasmon oscillations propagating along the chain of prolate spheroids with small separations and with the longitudinal dipole polarization, despite continuous character of the cord---this is pictorially sketched in Fig. \ref{axon-p} (right panel). To estimate characteristics of such collective dipole wave-type excitation along the cord sectors one can use the model of the chain of spherical electrolyte systems with suitably  diluted ion concentration (resulted in the same total ion number as in the cord fragments) and with decreased resonance frequency as for the longitudinal polarization  in the case of  highly prolate spheroid (or elongated cylinder). The model  allows quantitative estimation of relevant propagation characteristics  and the verification whether PP dynamics might fit to the observed features of the saltatory conduction of myelinated axon, the nature of which  being  mysterious and obscure as of yet.

PP wave-packet triggers at Ranvier nodes the release of ions through the nonmyelinated  membrane and the forming of the action potential. Relatively low (but beyond the ignition threshold) polarization induced by the PP initiates at a Ranvier node  opening of the $Na^+$ and $K^+$ inter-membrane ion channels, which results in a characteristic larger activation signal due to the transfer of ions through the opened gates caused by the difference in their concentrations on either sides of the membrane. The entire cycle requires a few milliseconds, but the initial increase in polarization caused by the rapid opening of the $Na^+$ channel occurs on the scale of a single millisecond. Because the myelin layer wrapped  by the Schwann cell prohibits cross-membrane ion transfer, the local polarization/depolarization of the internal cytoplasm of the axon occurs only in the Ranvier nodes, thus strengthening the dipole formation in the axon fragment wrapped  by the Schwann cell resulting in wave-type collective  dipole oscillations in the whole chain. The PP mode that traverses the axon structure consisting of periodically polarized electrolyte segments wrapped with the myelin sheath can cause the one-by-one individual ignitions of the Ranvier node sequence along the axon. This triggering role of the PP thus elucidates how the action signal jumps between neighboring active Ranvier nodes.

 The PP scenario in an ionic chain might share certain features in common with the ability of nerve systems to achieve very efficient and energetically economical electric signaling despite the rather poor ordinary conductivity of axons (the cable theory predicts velocity of signal by two orders lower than the saltatory conduction velocity). The  PP kinetics in the periodic structure of the axon may be thus a reasonable explanation for high efficiency of  nerve signaling by providing quick signal propagation despite  low ordinary conductivity and with a practically unlimited range and stable PP amplitude when energy is permanently supplied to cover  relatively small PP Joule losses (lower than for ordinary currents). Moreover, PPs do not irradiate any energy. The energy supply to cover Ohmic losses is provided residually   by the ATP/ADP mechanism in neuron cell, which energetically contributes to the signal-dependent opening and closing of $Na^+$ and $K^+$ channels in the Ranvier nodes and then to the restoration of the steady-state conditions, i.e., to the active transport of ions across the membrane against the ionic concentration gradient. Additionally, the coincidence of the micrometer  scale of the axon's periodic structure of Schwann cells (of approximately 100 $\mu$m in length) with the typical requirements for the size of ionic chains supports the suggested PP concept as the explanation of the transduction of the action potential along the periodically myelinated  axon in agreement with the tentative quantitative fitting.

It must also be emphasized that PPs do not interact with external electro-magnetic waves, or, equivalently, with photons (even with adjusted frequency---very small for PP in axons), which is a consequence of the large difference in group velocity between the PP and the photon velocity ($c/\sqrt{\varepsilon}$). The resulting large discrepancy in wavelength between a photon and a PP of the same energy prohibits the mutual transformation of the two types of excitation, PPs and cold photons, because of momentum conservation constraints. Therefore, PP signaling by means of collective wave-type dipole oscillations along a chain can be neither detected nor perturbed by external electro-magnetic radiation. This also well fits with neural signaling properties  in the peripheral nervous system  and in  white myelinated matter in the central system. 
It is also worth noting that surface plasmon frequencies are independent of temperature, although not independent for volume plasmons, as shown in Ref.  \citenum{jacak2014a}. However, the temperature influences the mean velocity of the ions, $\mathsf{v}=\sqrt{\frac{3kT}{m}}$, thereby causing an enhancement in Ohmic losses with increasing temperature (cf. Eq. (\ref{form})), which, in turn, strengthens the PP damping. Hence, at higher temperatures, higher external energy supplementation is required to maintain the same long-range propagation of PPs with constant amplitude. This property   is also consistent with experimental observations. Note, however, that energy costly is an active ion transport through the membrane to recover the steady state of each of Ranvier nodes after spike formation. Only residual part of this energy is sufficient to eventually maintain undamped PP propagation, which can be regarded in the scheme of the steady solution for forced and damped oscillator applied here to PP dynamics \cite{jacak2015} similarly as in the metallic chain activated by coupled  quantum dot system   \cite{andrianov2012}. 

Note  that the energy supply from the  consecutive excited Ranvier nodes continuously covering  PP Ohmic losses   results also in a convenient narrowing of the PP $k$-wave packet shifting it toward the long-wave limit (small $k$), which follows from the Fourier picture of the Dirac delta (the larger number of the chain segments contributes to the excitation of the PP, the narrower PP $k$-wave packet is formed). This favorable property allows for the precise ignition of sequent Ranvier nodes by unlimitedly propagating relatively narrow PP wave packet. 

Advantages of PP kinetics in agreement with observed myelinated neuron signaling properties can be listed as follows:
\begin{itemize}
\item{PPs cannot be neither perturbed nor detected by e-m waves (EEG signals are related to grey matter activity with ordinary conduction [according to the cable theory] and producing weak e-m field, whereas myelinated white matter does not produce EEG signal).}
\item{Immunity with respect to e-m perturbation is convenient in long links in  myelinated neurons of peripheral nervous system
 (motorics and sensing cannot be  disturbed by e-m field).}
\item{All the e-m field of PPs is compressed along the tight tunnel around the axon cord.} 
\item{The myelin sheath must be sufficiently thick not for insulation but for creation of dielectric tunnel in surrounding inter-cellular electrolyte to allow PP formation (reducing in myelin sheath thickness perturbs formation of PPs or lowers PP group velocity, resulting in  MS).}
\item{Increase of temperature causes enhancement of Ohmic losses but does not reduce the PP group velocity  (however, more energy ADP/ATP, residual to action potential spike formation, is needed to compensate the temperature increase of  Ohmic losses).}
\item{Any radiative losses do not occur  for PPs; if PP is residually aided by spike formation (renovation)  at consecutive Ranvier nodes, the arbitrary long range PP kinetics is possible with constant amplitude of the dipole signal}.
\item{The PP group velocity increases with the cross-section of the axon cord}.
\item{The PP group velocity for real neuron electrolyte and size parameters fits to observed velocity of saltatory conduction}.
\item{Wave type of PP kinetics explains  maintenance of the signal transduction despite of small breaks  and gaps in neuron cord or damage of some Ranvier nodes (these observations  were impossible to be explained either within the cable theory or by compression-soliton model).}
\item{One-direction kinetics if started from one end of the axon but two-direction kinetics if PP is ignited from some central point of an axon}.
\end{itemize}

\section{Summary} We proposed the nonlocal--topology type of braid group approach to information processing in brain taking into account the specific and different electricity of the gray and white matter of neurons. The web of neuron filaments in gray matter would serve to encode an information in entanglement of filaments selected by activation of some synapses. The method of e-m resonance identification of information stored in braided segments of neuron web is suggested and discussed. This method favors the gray matter with ordinary ion currents of diffusive type (upon the cable model) but excludes the myelinated white matter with wave-type signaling. The saltatory conduction in myelinated axons in the white matter  is not active with respect to e-m resonance. We discuss this in detail and propose the model of ionic plasmon-polariton propagation to explain the saltatory conduction. This wave type propagation of the action potential is actually passive with respect to e-m responce and cannot contribute to braid information processing in the gray matter, though conveniently fulfills communication role in the peripheral and central nervous systems. Plasmon-polariton model  leads to an explanation of the efficient, very fast  and long range so-called saltatory conduction in myelinated axons in peripheral neural systems and in white matter of the brain and spinal cord, where the velocity  and energy-efficiency of communication is of the most significance.  The coincidence with observations  supports the propriety of the proposed new  model for the saltatory conduction of action potential, which, on the other hand,  complementary fits to the different role and functioning of the gray matter. Both myelinated and non-myelinated structures  share thus  between them the different but complementary  roles in the whole neuron system.

\begin{acknowledgments}
Supported by the NCN project no. 2011/02/A/ST3/00116.
\end{acknowledgments}

%\bibliographystyle{spiebib}

%\bibliography{plasmons,abbr,plasmons1,pl,plazmony,literatura}

\end{article}

\begin{figure*}
\centering
\scalebox{0.65}{\includegraphics{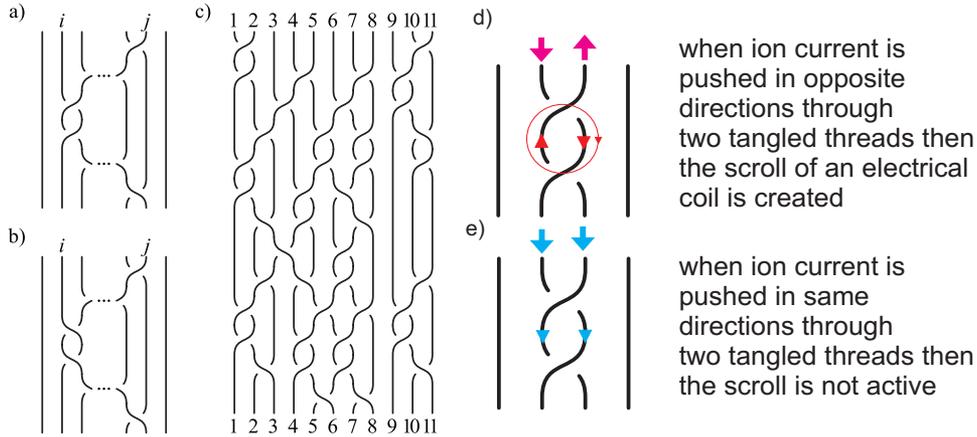}}
\caption{\label{coil} Generator $l_{ij}$ (a) and its inverse  $l^{-1}_{ij}$ b) of the pure braid group; an example of $N=11$ pure braid c); the scheme of creating of an e-m active d) (nonactive e)) scroll of coil by entanglement of threads in a pure braid. }
\end{figure*}

\begin{figure*}
\centering
\scalebox{1.2}{\includegraphics{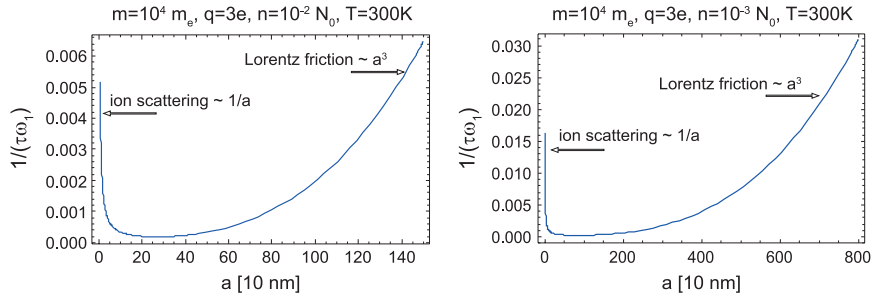}}
\caption{\label{lorentz45} The cross-over in the system size-dependence of the surface plasmon damping rate for $T=300$ K, $m=10^4m_e$, $q=3e$, $n=10^{-2}N_0$ (where $N_0$ is the concentration of one molar electrolyte) (left) and for $n=10^{-3}N_0$ (right); scattering losses, $\sim\frac{1}{a}$, are of lowering importance with growth of the electrolyte sphere size ($a$ is the sphere radius), whereas the irradiation losses dominate at larger size, as they are $\sim a^3$ }
\end{figure*}

\begin{figure*}
\centering
\scalebox{1.2}{\includegraphics{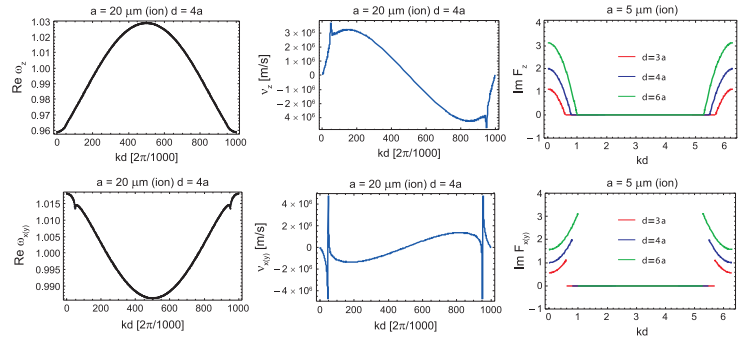}}
\caption{\label{figk1} The exact solutions for the self-frequencies of the longitudinally and transversely polarized modes of PPs in an ionic chain ($\omega$ in units of $\omega_1$) obtained by exactly solving Eq. (\ref{aaa1}) at 1000 points in the region $kd\in [0,2\pi$) (left) and the corresponding group velocities for both types of polarization (center); the functions $ImF_{z}(k;\omega=\omega_1)$ and $ImF_{x(y)}(k;\omega=\omega_1)$ for infinite chains of electrolyte spheres of radius $a$ at separations of $d=3a$, $4a$, and $6a$, the shift of the singularities toward band edges with lowering $d/a$ is noticeable (right)}
\end{figure*}

  \begin{figure*}
\centering
\scalebox{1.2}{\includegraphics{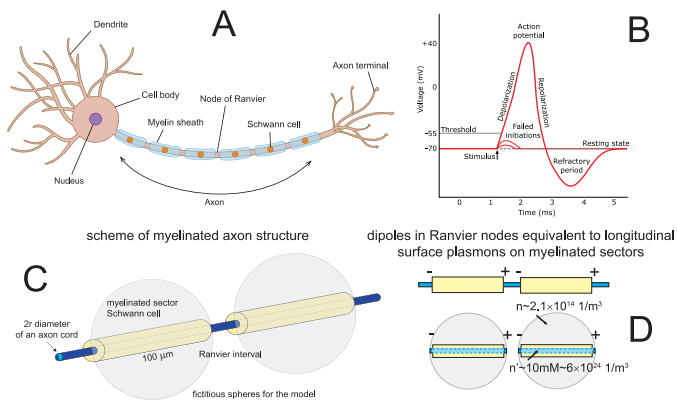}}
\caption{\label{axon-p} Schematic illustration of a long  axon with a chain of  periodically repeated myelinated  sectors  of approximately 100 $\mu$m in length separated by unmyelinated Ranvier nodes, corresponding to a number of segments of order 10 000 per 1 m of axon length (A); time-pattern of the action potential forming on a Ranvier node (B); periodic fragments of a myelinated  axon with a fictitious  periodic chain of spherical ionic systems  proposed as an effective model (C);  the equivalence of polarized Ranvier nodes with longitudinal surface plasmons on myelinated sectors  (effective concentration $n$ of ions in the auxiliary sphere corresponding to the actual ion concentration $n'$) (D)}
\end{figure*}

 \begin{figure*}
\centering
\scalebox{1.2}{\includegraphics{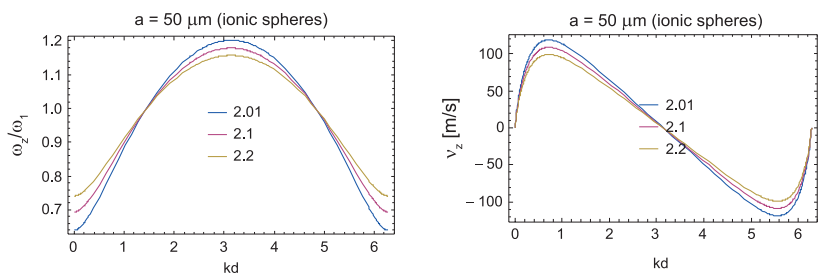}}
\caption{\label{axon-100} Solutions for the self-frequencies and group velocities of the longitudinal  mode of a PP in the model ionic chain; $\omega$ is presented in units of $\omega_1$, here $\omega_1=4 \times 10^6 $ 1/s, for a chain of spheres of radius $a=50\;\mu$m and a separation of $d/a=2.01$, $2.1$, or $2.2$  for an equivalent ion concentration in the inner ionic cord of the axon of $n'\sim 10$ mM}
\end{figure*}

  \begin{figure*}
\centering
\scalebox{1.2}{\includegraphics{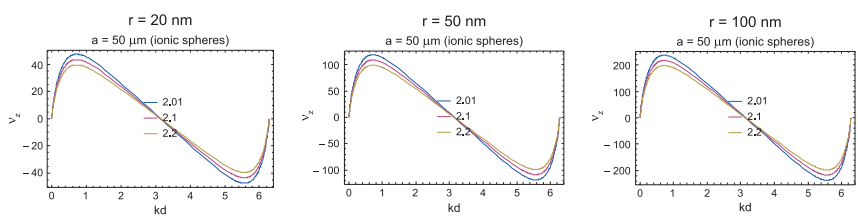}}
\caption{\label{axon-predkosc} Comparison of the group velocities, in units of m/s, of the longitudinal PP mode with respect to the wave vector $k\in[0,2\pi/d)$ within the axon model for a Schwann cell myelinated sectors with a length of 100 $\mu$m,  
Ranvier separations of 0.5 $\mu$m, 5 $\mu$m, and 10 $\mu$m (represented by $d/a=2.01$, $2.1$, and $2.2$ in the figure, respectively) and  for the axon cord radii of $r=20$, $50$, and $100$ nm }
\end{figure*}

  \begin{figure*}
\centering
\scalebox{1.2}{\includegraphics{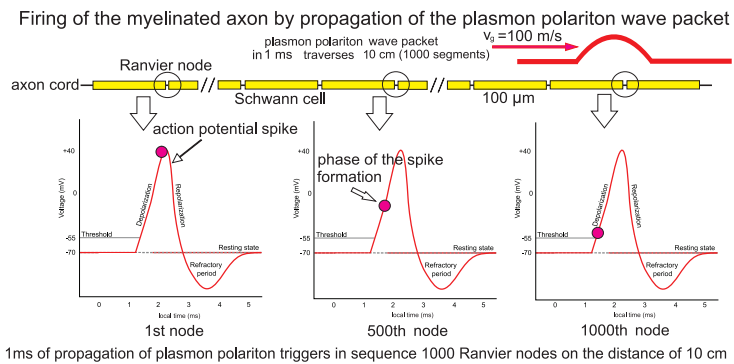}}
\caption{\label{fig666666} Schematic presentation of the firing of a nerve cell triggered by the propagation of the PP wave packet along the periodic structure of the myelinated axon; for group velocity of the PP wave packet of order of 100 m/s within 1 ms the packet traverses 10 cm distance and initiates one by one the forming of the action potential on 1000 in sequence Ranvier nodes being thus in various time phases as indicated by red dots for exemplary nodes.}
\end{figure*}

\end{document}